\documentclass[sigconf]{acmart}
\AtBeginDocument{%
  }

\copyrightyear{2025}
\acmYear{2025}
\setcopyright{cc}
\setcctype{by}
\acmConference[KDD '25]{Proceedings of the 31st ACM SIGKDD Conference on Knowledge Discovery and Data Mining V.2}{August 3--7, 2025}{Toronto, ON, Canada}
\acmBooktitle{Proceedings of the 31st ACM SIGKDD Conference on Knowledge Discovery and Data Mining V.2 (KDD '25), August 3--7, 2025, Toronto, ON, Canada}
\acmDOI{10.1145/3711896.3736979}
\acmISBN{979-8-4007-1454-2/2025/08}

\usepackage{listings}
\usepackage{titletoc}
\usepackage{graphicx}
\usepackage[noabbrev,capitalise]{cleveref}

\usepackage{amssymb}
\usepackage{multicol}
\usepackage{multirow}
\usepackage{xspace}
\usepackage{makecell}
\usepackage{bm}
\usepackage{subfigure}

\newcommand{\ie}{\emph{i.e.,}\xspace}
\newcommand{\eg}{\emph{e.g.,}\xspace}

\newcommand{\wrt}{w.r.t.\xspace}
\newcommand{\ignore}[1]{}
\newcommand{\paratitle}[1]{\vspace{1.5ex}\noindent\textbf{#1}}

\newcommand{\mymodel}{RPG}

\begin{document}

\title{Generating Long Semantic IDs in Parallel for Recommendation}

\author{Yupeng Hou}
\email{yphou@ucsd.edu}
\affiliation{%
  \institution{University of California, San Diego}
  \city{La Jolla}
  \country{United States}
}

\author{Jiacheng Li}
\email{jiachengli@meta.com}
\affiliation{%
  \institution{Meta AI}
  \city{Sunnyvale}
  \country{United States}
}

\author{Ashley Shin}
\email{ashleyshin@ucsd.edu}
\affiliation{%
  \institution{University of California, San Diego}
  \city{La Jolla}
  \country{United States}
}

\author{Jinsung Jeon}
\email{jij014@ucsd.edu}
\affiliation{%
  \institution{University of California, San Diego}
  \city{La Jolla}
  \country{United States}
}

\author{Abhishek Santhanam}
\email{absanthanam@ucsd.edu}
\affiliation{%
  \institution{University of California, San Diego}
  \city{La Jolla}
  \country{United States}
}

\author{Wei Shao}
\email{weis@meta.com}
\affiliation{%
  \institution{Meta AI}
  \city{Sunnyvale}
  \country{United States}
}

\author{Kaveh Hassani}
\email{kavehhassani@meta.com}
\affiliation{%
  \institution{Meta AI}
  \city{Toronto}
  \country{Canada}
}

\author{Ning Yao}
\email{nyao@meta.com}
\affiliation{%
  \institution{Meta AI}
  \city{Sunnyvale}
  \country{United States}
}

\author{Julian McAuley}
\email{jmcauley@ucsd.edu}
\affiliation{%
  \institution{University of California, San Diego}
  \city{La Jolla}
  \country{United States}
}

\renewcommand{\shortauthors}{Yupeng Hou et al.}

\begin{abstract}
Semantic ID-based recommendation models tokenize each item into a small number of discrete tokens that preserve specific semantics, leading to better performance, scalability, and memory efficiency. While recent models adopt a generative approach, they often suffer from inefficient inference due to the reliance on resource-intensive beam search and multiple forward passes through the neural sequence model. As a result, the length of semantic IDs is typically restricted (\eg to just $4$ tokens), limiting their expressiveness.
To address these challenges, we propose \mymodel, a lightweight framework for semantic ID-based recommendation. The key idea is to produce unordered, long semantic IDs, allowing the model to predict all tokens in parallel.
We train the model to predict each token independently using a multi-token prediction loss, directly integrating semantics into the learning objective. During inference, we construct a graph connecting similar semantic IDs and guide decoding to avoid generating invalid IDs.
Experiments show that scaling up semantic ID length to $64$ enables RPG to outperform generative baselines by an average of 12.6\% on the NDCG@10, while also improving inference efficiency.
Code is available at: \textcolor{blue}{\url{https://github.com/facebookresearch/RPG_KDD2025}}.
\end{abstract}

\ccsdesc[500]{Information systems~Recommender systems}

\keywords{Sequential Recommendation; Semantic ID}

\maketitle

\section{Introduction}

A semantic ID represents an item as a sequence of discrete tokens from a shared vocabulary~\cite{hou2023vqrec,rajput2023tiger,zhai2024hstu,zheng2024graph_semid}. Unlike traditional recommender systems that rely on unique item IDs~\cite{hidasi2016gru4rec,kang2018sasrec}, semantic IDs eliminate the need for large embedding tables that scale with the total number of items.
Early semantic ID-based recommendation models follow a retrieval paradigm, learning vector representations for items based on their semantic IDs and retrieving those closest to the user representation~\cite{hou2023vqrec,petrov2024recjpq}. Although effective, these models still require runtime memory proportional to the number of items.
More recently, a generative paradigm has emerged~\cite{geng2022p5,rajput2023tiger}, where models are trained to autoregressively generate the next semantic ID in a token-by-token manner.
These generative models make recommendations by decoding in the semantic ID space, resulting in runtime memory and latency independent of the number of items. Additionally, fine-grained semantics are explicitly integrated into the training objective by having the model generate sub-item tokens instead of entire items, resulting in strong performance.

Despite their promise, semantic ID-based generative models introduce significant \textbf{latency issue in model inference}. Compared to retrieval-based methods, where recommendations involve a single-step decoding process (\ie each token represents one item, and the next item is recommended by decoding the next token), generative models require multi-step decoding.  
For example, TIGER~\cite{rajput2023tiger} autoregressively generates the next item's semantic IDs in a token-by-token manner. In this multi-step decoding process, each step requires at least one batch of sequence model forwarding. To generate a ranking list of size $K$, TIGER employs beam search~\cite{graves2012beamsearch}, maintaining the top $b$ beams (semantic ID prefixes) and forwarding them through the neural encoder at each step. Because of autoregressive token generation and beam search, generative recommendation models require significantly more neural network forward passes for a single recommendation.  

Moreover, the inference methods of existing generative recommendation models \textbf{limit the expressiveness of semantic IDs}. To ensure token-by-token autoregressive generation runs with reasonable latency, these models typically tokenize each item into a short sequence of semantic IDs. For example, TIGER uses $4$ tokens per item, resulting in a decoding process of $4$ steps per recommendation. However, representing an item with so few discrete tokens may be insufficient, especially as item features carry richer semantics. Retrieval-based methods like VQ-Rec~\cite{hou2023vqrec} use longer semantic ID sequences (\eg $32$ tokens per item), but applying such long sequences in generative models is impractical. Running beam search for $32$ steps would be computationally prohibitive.

In this work, we aim to develop a semantic ID-based recommendation model that offers both efficient inference and expressive semantic IDs.
Instead of generating semantic IDs token by token, we propose generating all tokens of the next item's semantic ID in parallel. This approach allows us to decode the entire semantic ID in a single step. Since the number of decoding steps is independent of the semantic ID length, we can design expressive and long semantic IDs. However, applying this idea is non-trivial due to the following challenges:
\textbf{(1)~Sparse decoding space.} Generating all tokens in parallel removes sequential dependencies between tokens. However, it also means that predicting one token is independent of the predictions of other tokens. This lack of awareness makes it extremely difficult to map the predicted token combination to an existing semantic ID from the item pool. For example, if each semantic ID consists of $32$ tokens, with each token selected from a codebook of $256$, the total number of possible semantic IDs is $256^{32}\simeq 10^{77}$. Even if there are billions of items, the valid semantic IDs remain sparse compared to this vast decoding space ($10^{9} \ll 10^{77}$).
\textbf{(2)~Efficient unordered decoding.} In recommendation tasks, we typically generate a top-$K$ recommendation list for each input instance. Naively enumerating all candidate items would require runtime memory similar to retrieval-based methods, negating the benefits of generative approaches. Existing decoding methods, such as beam search, assume that tokens within a semantic ID follow a specific order. However, this assumption is unnecessary when generating tokens in parallel. Therefore, a well-designed semantic ID decoding method is needed to handle unordered tokens efficiently.

To this end, we propose \mymodel, which stands for \underline{R}ecommendation with \underline{P}arallel semantic ID \underline{G}eneration. To align with parallel decoding, we tokenize each item into an unordered long semantic ID (up to $64$ tokens per item) using optimized product quantization~\cite{jegou2011pq,ge2014opq}. To enable the model to generate semantic IDs in parallel, we train it with a multi-token prediction (MTP) objective~\cite{gloeckle2024mtp} instead of the conventional next-token prediction objective~\cite{rajput2023tiger}. By applying MTP, sub-item semantics are explicitly incorporated into the training objective, similar to next-token prediction.
Given MTP as the training objective, the prediction score for a candidate semantic ID is computed by summing the log probabilities of the ground-truth tokens across all digits. A key observation is that when two semantic IDs differ in only a few digits, their prediction scores should remain similar. Inspired by the widely used constrained beam search~\cite{hokamp2017constraned_beam_search}, which constructs a tree structure linking semantic prefixes that differ by only one digit, we propose a graph-constrained decoding method for unordered semantic IDs. This method connects similar semantic IDs (which have close prediction scores) by an edge in the graph structure.
The decoding process follows an iterative graph propagation approach. Starting with a randomly initialized beam of semantic IDs, we propagate on the graph to identify semantic IDs that are close to the existing beam. We retain those with higher prediction scores and repeat the process for a predefined number of iterations to obtain the final top-$K$ recommendations.

We show that the runtime memory and time complexities of RPG's inference process are independent of the number of candidate items. Extensive experiments on three public benchmarks and a larger public dataset demonstrate that RPG benefits from expressive long semantic IDs and parallel generation, achieving the overall best performance among the compared baselines while being more efficient than existing semantic ID-based generative methods.

\begin{figure*}[t]
  \centering
  \includegraphics[width=0.95\textwidth]{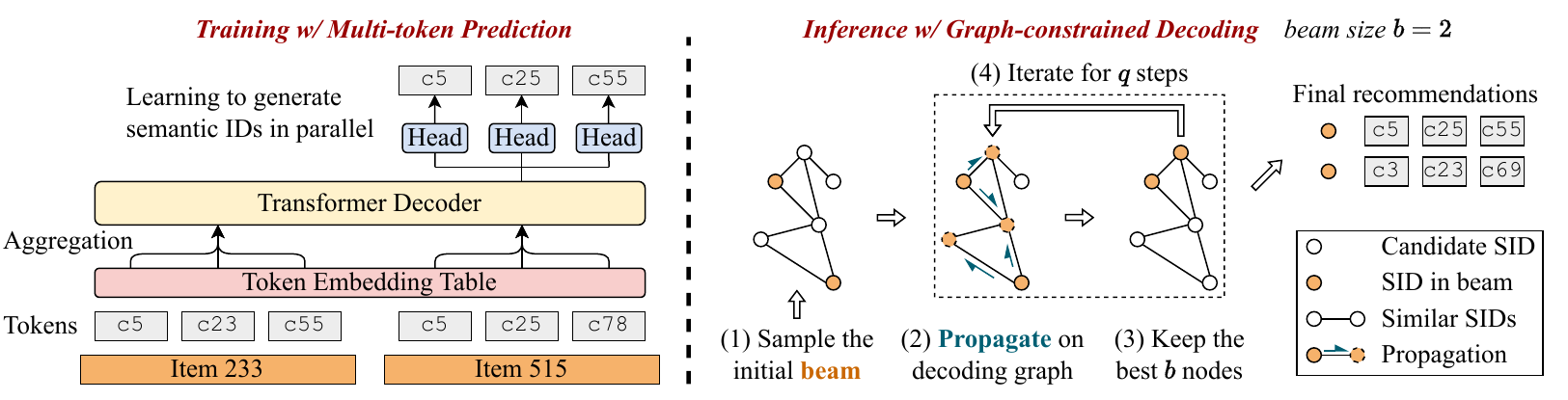}
  \caption{The overall framework of \mymodel. }
  \label{fig:model}
\end{figure*}

\section{Method}

In this section, we introduce \mymodel, a recommendation model that generates each token of the next semantic ID in parallel. We formulate this task as a sequential recommendation problem.
Given a user, \mymodel\ takes a sequence of previously interacted items, $s = \{i_1, i_2, \ldots, i_{t-1}\}$, as input and predicts the next item $i_t$. 
First, we describe how \mymodel\ tokenizes and represents each item using a long semantic ID (\Cref{sec:item-embedding}). Next, we explain how \mymodel\ learns to generate semantic IDs in parallel by optimizing the multi-token prediction objective (\Cref{sec:dual-level-training}). We then introduce graph-constrained decoding, which efficiently decodes unordered long semantic IDs during inference (\Cref{sec:decoding}). Finally, we discuss the connections between \mymodel\ and both retrieval-based and generative models. \Cref{fig:model} illustrates the overall framework of the proposed approach.

\subsection{Long Semantic ID-based Item Representation}\label{sec:item-embedding}

In recommender systems, each item is typically associated with multiple content features (\eg text descriptions and images). A common pipeline for item representation first encodes raw features into dense vector representations using a semantic encoder (\eg pretrained language models~\cite{ni2022sentence_t5,ni2022gtr}). Then, these vectors are tokenized into discrete tokens.
Most existing semantic ID-based generative approaches use relatively short semantic IDs, typically with only four digits, due to efficiency concerns. In this work, we aim to design a model that generates each token in a semantic ID in parallel, enabling the use of longer semantic IDs without significantly impacting efficiency.
In this section, we describe how we convert raw features into semantic IDs and how we represent each item using long semantic IDs.

\subsubsection{Long Semantic ID Construction} \label{subsubsec:sid_construction}

After obtaining semantic representations for each item~\cite{rajput2023tiger,zheng2024lcrec}, we adopt product quantization instead of the commonly used residual quantization-based methods (\eg RQ-VAE~\cite{zeghidour2022rqvae,rajput2023tiger}). There are two key reasons for this choice: (1) predicting tokens in semantic IDs in parallel eliminates the need for sequential dependencies in RQ-based methods; (2) existing studies suggest that RQ-based semantic IDs suffer from unbalanced information distribution among tokens and poor scalability~\cite{zhang2024moc}.  
Concretely, following~\citet{hou2023vqrec}, we use optimized product quantization (OPQ)~\cite{jegou2011pq,ge2014opq} to produce a tuple of $m$ tokens $(c_1, \ldots, c_{m})$ for each item $i$, where \( c_1 \in \mathcal{C}^{(1)}=\{1,\ldots,M\}, c_2\in\mathcal{C}^{(2)}=\{M+1, \ldots, 2M\} \), and so on. Here, $M$ represents the codebook size for each digit of the semantic ID. Each token in the tuple is drawn from a different codebook, with each codebook corresponding to the semantic information of a subvector of the original semantic representation.  
In this work, we set the semantic ID length $m$ to a maximum of 64, which is significantly longer than the typical length of 4 used in generative recommendation models.

\subsubsection{Semantic ID Embedding Aggregation} \label{subsubsec:emb_aggr}

After tokenizing each item into a semantic ID of length \( m \), we consider how to represent each item in the input sequence. For each codebook, we assign a learnable token embedding table \( \bm{E}_j \in \mathbb{R}^{M \times d} \). This allows us to obtain the corresponding token embeddings \( (\bm{e}_{1, c_1}, \ldots, \bm{e}_{m, c_{m}}) \) for item \( i \).  
Given that semantic IDs can be as long as 64 in this work, directly concatenating them may make the input token sequence too long, leading to efficiency and context length challenges~\cite{rajput2023tiger}. To address this, we adopt an item-level input representation for sequence modeling. Specifically, we aggregate the token embeddings into a single item representation, \( \bm{v}_i \in \mathbb{R}^d \), defined as  $\bm{v}_i = \operatorname{Aggr}(\bm{e}_{1,c_1}, \ldots, \bm{e}_{m,c_{m}})$, where \( \operatorname{Aggr}(\cdot) \) is an aggregation function. In practice, we can use mean or max pooling to aggregate the token embeddings efficiently~\cite{hou2023vqrec}.

\subsection{Learning to Generate Semantic IDs in Parallel}\label{sec:dual-level-training}

Given the input item representation sequence \(\{\bm{v}_1, \bm{v}_2, \ldots, \bm{v}_{t-1}\}\) and the target semantic ID \((c_{t,1}, c_{t,2}, \ldots, c_{t,m})\), we first introduce the learning objective that allows the model to generate tokens in parallel (\Cref{subsubsec:mtp}). Then, we present an efficient method for logit calculation in models optimized with this objective.

\subsubsection{Multi-token Prediction Objective}\label{subsubsec:mtp}

The key idea behind an efficient semantic ID-based recommendation model is to predict all tokens in the next semantic IDs in parallel. To achieve this, we draw inspiration from the multi-token prediction (MTP) objective~\cite{gloeckle2024mtp}. First, we encode the item representation sequence using Transformer decoders, obtaining a sequence representation \(\bm{s} \in \mathbb{R}^d\).  
We then formulate the problem as training the model to fit the probability \(\mathbb{P}(c_{t,1}, \ldots, c_{t,m}|s)\), which represents the likelihood that a given token combination appears in the next semantic IDs among all possible combinations. Since each token produced by OPQ encodes a separate subvector of the target item's semantic representation, we assume that tokens in the tuple are conditionally independent given the input sequence \(s\). This allows us to factorize the probability as \(\prod_{j=1}^{m} \mathbb{P}^{(j)}(c_{t,j}|s)\), where \(\mathbb{P}^{(j)}(c_{t,j}|s)\) is the probability of the \(j\)-th token being in the next semantic ID.  
Formally, the MTP loss is defined as:  
\begin{align}
    \mathcal{L} &= -\sum_{j=1}^{m} \log \mathbb{P}^{(j)}(c_{t,j}|s) = - \sum_{j=1}^m\log \frac{\exp(\bm{e}_{c_{t,j}}^\top \cdot \operatorname{g}_j(\bm{s}) / \tau)}{\sum_{c \in \mathcal{C}^{(j)}} \exp(\bm{e}_{c}^\top \cdot \operatorname{g}_j(\bm{s}) / \tau)},\label{eq:token-level-loss}
\end{align}  
where 
$\tau$ is the temperature hyperparameter, \(\operatorname{g}_j(\cdot)\) is a projection head (specifically, an MLP module~\cite{chen2020simclr,cai2024medusa}) that maps sequence representations to the space of tokens in codebook \(\mathcal{C}^{(j)}\).

Note that the MTP loss functions differently from the widely used cross-entropy loss defined in the item pool~\cite{kang2018sasrec,hou2023vqrec}. MTP loss optimizes the probability of the target semantic ID in the token space, \ie, against all possible token combinations. In contrast, traditional item-based cross-entropy loss optimizes the probability of the ground-truth item ID in the item space, \ie, against all other items.
Additionally, similar to the next-token prediction loss used in generative recommendation models~\cite{rajput2023tiger,zhai2024hstu}, MTP directly integrates tokens in semantic IDs into the learning objective, enabling fine-grained semantic learning.

\subsubsection{Efficient Logit Calculation}\label{subsubsec:logit}

Given a model optimized with MTP loss, we propose an efficient logit calculation method. During inference, we compute \(\mathcal{L}\) in~\Cref{eq:token-level-loss} as the logits. A straightforward approach is to calculate logits for each candidate item individually by enumerating token embeddings, computing logits for each token, and summing them. This results in a time complexity of \(O(Nmd)\) for \(N\) candidate items. However, this method involves redundant dot product calculations between sequence representations and token embeddings. Popular tokens may undergo multiple dot product operations with the sequence representation.  
To address this inefficiency, we first cache the dot product between the sequence representation and all token embeddings:  
\begin{equation}
\bm{p}^{(j)} = \operatorname{softmax}(\bm{E}_j\cdot \operatorname{g}_j(\bm{s}) / \tau) \in \mathbb{R}^{M},\label{eqn:cache}
\end{equation}
where \(M\) is the codebook size, independent of the number of candidate items. Then, for a candidate semantic ID \((c_1, \ldots, c_m)\), the logits can be computed as:
\begin{equation}
    \sum_{j=1}^{m} \log \bm{p}^{(j)}_{c_j}, \label{eqn:fast_logits}
\end{equation}
achieving a reduced time complexity of \(O(Mmd + Nm)\). This cached version is more efficient when \(N > \frac{d}{d-1}M\). The choice of method depends on the serving scenario, and people can decide which inference approach works best for their needs.

\subsection{Next Semantic ID Decoding with Graph Constraints}\label{sec:decoding}

For a model trained with the MTP objective, we aim to develop a decoding approach that is: (1) more efficient than naively enumerating all candidate items or applying beam search, and (2) more likely to generate valid semantic IDs rather than illegal token combinations.

\begin{figure}[!t]
    \begin{center}
    \includegraphics[width=0.95\columnwidth]{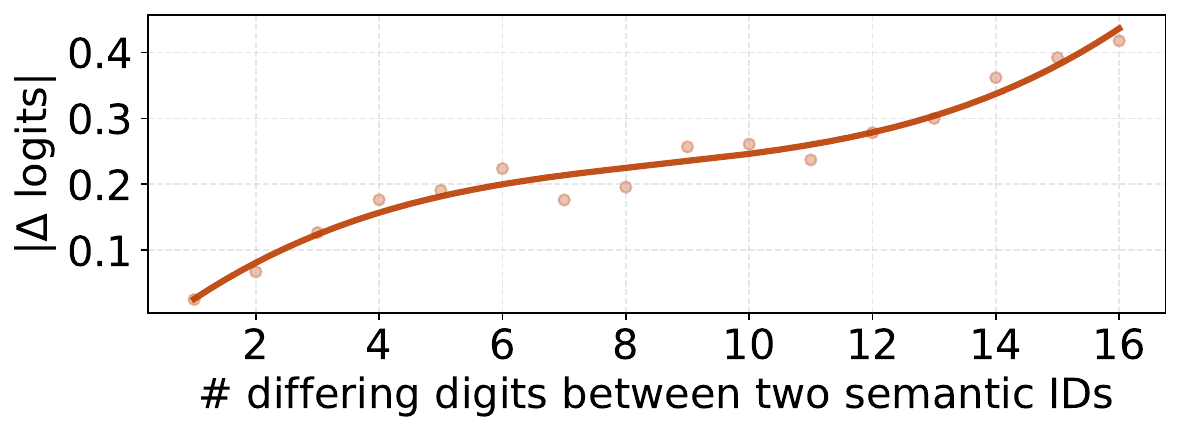}
    \caption{The relationship between the number of differing digits in two semantic IDs and the absolute difference in their predicted logits on the ``Sports'' dataset. As the number of differing digits increases, the model predicts increasingly different logits.}
    \label{fig:case}
    \end{center}
    \vskip -0.5cm
\end{figure}

\paratitle{Observation.} The key insight lies in the unordered nature of PQ-based semantic IDs and logit calculation. Specifically, the logit of a candidate item is computed by summing the logits of each token in the corresponding semantic ID (\Cref{eqn:fast_logits}). Consequently, when two semantic IDs differ by only a few digits, their logits should remain similar, as many token logits are identical. This observation is also empirically verified in~\Cref{fig:case}.

\paratitle{Overview.} Inspired by constrained beam search~\cite{hokamp2017constraned_beam_search}, which uses a tree structure to connect similar semantic prefixes, we propose a graph-constrained semantic ID decoding method. This method links similar semantic IDs through a graph structure to guide the decoding process. The main idea is to start by sampling a random set of semantic IDs, propagate through the graph to explore new semantic IDs with slightly higher or lower logits, retain those with the best logits, and repeat this process for a fixed number of steps.

\subsubsection{Building Graph to Connect Similar Semantic IDs}

After training a model, we construct the \emph{decoding graph}, which connects semantically similar semantic IDs. A node represents a valid semantic ID from the item pool. For any two nodes, \((c_{1,1}, \ldots, c_{1,m})\) and \((c_{2,1}, \ldots, c_{2,m})\), we compute their semantic similarity as \(\sum_{j=1}^m \bm{e}^{\top}_{j,c_{1,j}} \cdot \bm{e}_{j,c_{2,j}}\). These similarity scores serve as edge weights in the graph.  
Since the initial graph is dense, it is inefficient and impractical. To address this, we sparsify it by retaining only the top \(k\) edges with the highest similarities for each node. Once built, the graph can be used throughout model inference until the model is retrained or the items are re-tokenized.

\subsubsection{Decoding with Iterative Graph Propagation}

Given an input sequence, we first cache the token logits following~\Cref{eqn:cache}.

\textbf{(1) Sampling an initial beam.} Decoding begins by sampling $b$ semantic IDs from the item pool. We use the term ``beam'' by analogy with beam search, where $b$ denotes the beam size.

\textbf{(2) Propagating on the decoding graph.} Similar to constrained beam search, the next step is to explore a larger set of semantic prefixes (or semantic IDs) and retain the best ones. Since neighboring nodes in the graph tend to have similar logits, we expand the beam by propagating through the graph, including all $b \times k$ neighbors of the current beam into a temporary set.

\textbf{(3) Keeping the best $b$ nodes.} After expanding the beam, we compute their logits using the cached logits from~\Cref{eqn:fast_logits}. The $b$ nodes with the highest logits are then retained as the next beam.

\textbf{(4) Iterating for $q$ steps.} We repeat steps (2) and (3) for $q$ iterations to obtain the final results. Since each node has an edge to itself, this iterative beam update ensures that the average logits never decrease. In the worst case, if the beam remains unchanged for two consecutive steps, the logits will stay the same. The semantic IDs in the final beam are recommended as top $K$ predictions.

\subsubsection{Complexity Analysis}\label{subsubsec:complexity}

The decoding method has $q$ steps, with each step involving at most $O(bk)$ nodes. This results in a time complexity of $O(Mmd + bqkm)$ by replacing $O(N)$ with $O(bqk)$ in~\Cref{subsubsec:logit}, making it independent of the number of items. In practice, the graph propagation process can be further parallelized, \eg by using distributed computation techniques like MapReduce~\cite{dean2008mapreduce}.

\subsection{Discussion}

In this section, we discuss the differences and connections between \mymodel\ and both retrieval-based and generative models.

\subsubsection{Comparison of \mymodel\ and Retrieval-based Models}

Compared to retrieval-based models like VQ-Rec~\cite{hou2023vqrec}, \mymodel\ offers two advantages that are typically exclusive to generative models: (i) its inference time complexity is independent of the total number of items, and (ii) it explicitly integrates sub-item level semantics into the learning objective.
While the proposed graph-constrained decoding can be viewed as a form of approximate nearest neighbor (ANN) search, most existing ANN methods, such as those based on quantization codes~\cite{douze2024faiss}, still require enumerating all items, resulting in time complexity proportional to the total number of items.

\subsubsection{Comparison of \mymodel\ and Generative Models}

Compared to generative models like TIGER~\cite{rajput2023tiger}, \mymodel\ reduces the number of sequence encoder forward passes per input instance from approximately $O(bm)$ to just $O(1)$, greatly improving inference efficiency (\Cref{subsubsec:exp-efficiency}). Additionally, \mymodel\ benefits from a long semantic ID, which is empirically shown to be more expressive (\Cref{subsubsec:exp-expressive}).

\begin{table}[!t] %
  \small
  \centering
\caption{Statistics of the processed datasets. ``Avg.~$t$'' denotes the average number of interactions per input sequence.}
\label{tab:dataset}
\begin{tabular}{crrrr}
  \toprule
  \textbf{Datasets} & \textbf{\#Users} & \textbf{\#Items} & \textbf{\#Interactions} & \textbf{Avg.~$t$}\\
  \midrule
  \textbf{Sports}  & 18,357            & 35,598           & 260,739            & 8.32 \\
  \textbf{Beauty}  & 22,363            & 12,101           & 176,139            & 8.87 \\
  \textbf{Toys}    & 19,412            & 11,924           & 148,185            & 8.63 \\
  \textbf{CDs}     & 75,258            & 64,443           & 1,022,334          & 14.58 \\
  \bottomrule
\end{tabular}
\end{table}

\begin{table*}[t!]
  \small
    \centering
    \caption{Performance comparison among baselines and \mymodel. The best performance score is denoted in \textbf{bold}. The second-best performance score is denoted in \underline{underline}. 
    \mymodel\ statistically outperforms the best baselines at a significance level of $p<0.05$ by paired t-test.
  }
    \label{tab:overall}
  \setlength{\tabcolsep}{1.2mm}{
    \begin{tabular}{@{}lcccccccccccccccc@{}}
      \toprule
      \multicolumn{1}{c}{\multirow{2.5}{*}{\textbf{Model}}} & \multicolumn{4}{c}{\textbf{Sports and Outdoors}} & \multicolumn{4}{c}{\textbf{Beauty}} & \multicolumn{4}{c}{\textbf{Toys and Games}} & \multicolumn{4}{c}{\textbf{CDs and Vinyl}} \\
      \cmidrule(l){2-5} \cmidrule(l){6-9}\cmidrule(l){10-13}\cmidrule(l){14-17}
      & R@5 & N@5 & R@10 & N@10 & R@5 & N@5 & R@10 & N@10 & R@5 & N@5 & R@10 & N@10 & R@5 & N@5 & R@10 & N@10 \\
      \midrule
      \midrule
      \multicolumn{17}{@{}c}{\textit{Item ID-based}} \\
      \midrule
      Caser & 0.0116 & 0.0072 & 0.0194 & 0.0097 & 0.0205 & 0.0131 & 0.0347 & 0.0176 & 0.0166 & 0.0107 & 0.0270 & 0.0141 & 0.0116 & 0.0073 & 0.0205 & 0.0101 \\
GRU4Rec & 0.0129 & 0.0086 & 0.0204 & 0.0110 & 0.0164 & 0.0099 & 0.0283 & 0.0137 & 0.0097 & 0.0059 & 0.0176 & 0.0084 & 0.0195 & 0.0120 & 0.0353 & 0.0171 \\
HGN & 0.0189 & 0.0120 & 0.0313 & 0.0159 & 0.0325 & 0.0206 & 0.0512 & 0.0266 & 0.0321 & 0.0221 & 0.0497 & 0.0277 & 0.0259 & 0.0153 & 0.0467 & 0.0220 \\
BERT4Rec & 0.0115 & 0.0075 & 0.0191 & 0.0099 & 0.0203 & 0.0124 & 0.0347 & 0.0170 & 0.0116 & 0.0071 & 0.0203 & 0.0099 & 0.0326 & 0.0201 & 0.0547 & 0.0271 \\
SASRec & 0.0233 & 0.0154 & 0.0350 & 0.0192 & 0.0387 & 0.0249 & 0.0605 & 0.0318 & 0.0463 & 0.0306 & 0.0675 & 0.0374 & 0.0351 & 0.0177 & 0.0619 & 0.0263 \\
FDSA & 0.0182 & 0.0122 & 0.0288 & 0.0156 & 0.0267 & 0.0163 & 0.0407 & 0.0208 & 0.0228 & 0.0140 & 0.0381 & 0.0189 & 0.0226 & 0.0137 & 0.0378 & 0.0186 \\
S$^3$-Rec & 0.0251 & 0.0161 & 0.0385 & 0.0204 & 0.0387 & 0.0244 & 0.0647 & 0.0327 & 0.0443 & 0.0294 & 0.0700 & 0.0376 & 0.0213 & 0.0130 & 0.0375 & 0.0182 \\
      \midrule
      \multicolumn{17}{@{}c}{\textit{Semantic ID-based}} \\
      \midrule
      RecJPQ & 0.0141 & 0.0076 & 0.0220 & 0.0102 & 0.0311 & 0.0167 & 0.0482 & 0.0222 & 0.0331 & 0.0182 & 0.0484 & 0.0231 & 0.0075 & 0.0046 & 0.0138 & 0.0066 \\
VQ-Rec & 0.0208 & 0.0144 & 0.0300 & 0.0173 & 0.0457 & 0.0317 & 0.0664 & 0.0383 & 0.0497 & 0.0346 & \underline{0.0737} & 0.0423 & 0.0352 & 0.0238 & 0.0520 & 0.0292 \\
TIGER & \underline{0.0264} & \underline{0.0181} & 0.0400 & \underline{0.0225} & 0.0454 & \underline{0.0321} & 0.0648 & 0.0384 & \underline{0.0521} & \underline{0.0371} & 0.0712 & \underline{0.0432} & \underline{0.0492} & \underline{0.0329} & \textbf{0.0748} & \underline{0.0411} \\
HSTU & 0.0258 & 0.0165 & \underline{0.0414} & 0.0215 & \underline{0.0469} & 0.0314 & \underline{0.0704} & \underline{0.0389} & 0.0433 & 0.0281 & 0.0669 & 0.0357 & 0.0417 & 0.0275 & 0.0638 & 0.0346 \\
      \midrule
      \textbf{\mymodel} & \textbf{0.0314} & \textbf{0.0216} & \textbf{0.0463} & \textbf{0.0263} & \textbf{0.0550} & \textbf{0.0381} & \textbf{0.0809} & \textbf{0.0464} & \textbf{0.0592} & \textbf{0.0401} & \textbf{0.0869} & \textbf{0.0490} & \textbf{0.0498} & \textbf{0.0338} & \underline{0.0735} & \textbf{0.0415} \\
      
      \bottomrule
    \end{tabular}
    }
  \end{table*}

\section{Experiments}\label{sec:exp}

In this section, we empirically demonstrate the effectiveness and efficiency of the proposed method \mymodel.

\subsection{Experimental Setup}

\textbf{Dataset.} We conduct experiments using four categories from the Amazon Reviews dataset~\cite{mcauley2015amazon}: ``Sports and Outdoors'' (\textbf{Sports}), ``Beauty'' (\textbf{Beauty}), ``Toys and Games'' (\textbf{Toys}), and ``CDs and Vinyl'' (\textbf{CDs}).  
The first three categories are commonly used for benchmarking semantic ID-based generative recommendation methods~\cite{rajput2023tiger,hua2023p5cid,jin2024lmindexer}.  
Additionally, we include ``CDs'', which has approximately four times more interactions than ``Sports'', to evaluate performance on larger datasets.  

Following previous studies~\cite{rajput2023tiger,hou2023vqrec,zhou2020s3rec}, we treat users' historical reviews as ``interactions'' and arrange them chronologically to form input interaction sequences, with earlier reviews appearing first.  
We adopt the widely used leave-last-out evaluation protocol~\cite{kang2018sasrec,zhao2022revisiting,rajput2023tiger}, where the last item in each sequence is reserved for testing and the second-to-last item for validation.  
\Cref{tab:dataset} presents the detailed statistics of the four datasets.  

\textbf{Baselines.}
We evaluate the performance of \mymodel\ by comparing it with the following item ID-based and semantic ID-based baselines.

  \noindent \hspace*{3mm} $\bullet$ Caser~\cite{tang2018caser} encodes the item ID sequence patterns using convolutional filters.\\
  \hspace*{3mm} $\bullet$ GRU4Rec~\cite{hidasi2016gru4rec} is a recurrent neural network (RNN)-based approach proposed for session-based recommendation.\\
  \hspace*{3mm} $\bullet$ HGN~\cite{ma2019hgn} applies the gating mechanism to RNN-based models.\\
  \hspace*{3mm} $\bullet$ BERT4Rec~\cite{sun2019bert4rec} uses a bidirectional Transformer encoder to model item ID sequences and is trained with a Cloze-style objective.\\
  \hspace*{3mm} $\bullet$ SASRec~\cite{kang2018sasrec} uses a self-attentive Transformer decoder to model item ID sequences by optimizing a binary cross-entropy objective.\\
  \hspace*{3mm} $\bullet$ FDSA~\cite{zhang2019fdsa} 
  processes the item ID sequences and item feature sequences separately through self-attention blocks.\\
  \hspace*{3mm} $\bullet$ S$^3$-Rec~\cite{zhou2020s3rec} is first pretrained by optimizing self-supervised objectives to capture the correlations between item features and item IDs. Then the model is fine-tuned for next-item prediction objective, using only item IDs.\\
  \hspace*{3mm} $\bullet$ VQRec~\cite{hou2023vqrec} tokenizes each item using product quantization into semantic IDs. The semantic ID embeddings are pooled together to represent each item.\\
  \hspace*{3mm} $\bullet$ TIGER~\cite{rajput2023tiger} tokenizes each item using RQ-VAE. The model is then trained to autoregressively generate the next semantic ID token by token and employs beam search for inference.\\
  \hspace*{3mm} $\bullet$ RecJPQ~\cite{petrov2024recjpq} applies joint product quantization to replace item embeddings with a concatenation of shared sub-embeddings.\\
  \hspace*{3mm} $\bullet$ HSTU~\cite{zhai2024hstu} discretizes raw item features into tokens, treating them as input tokens for generative recommendation. To ensure a fair comparison, we use the same $4$-digit OPQ-tokenized semantic IDs as the item tokens.

\textbf{Evaluation setting.} We use Recall$@K$ and NDCD$@K$ as metrics to evaluate the methods, where $K \in \{5, 10\}$, following~\citet{rajput2023tiger}. Model checkpoints with the best performance on the validation set are used for evaluation on the test set.

\textbf{Implementation details.} We use most of the baseline results from~\citet{rajput2023tiger}, while implementing other baselines mainly using RecBole~\cite{zhao2021recbole}. We carefully tune the baselines' hyperparameters following the original papers' suggestions. For model training, we train all implemented models for a maximum of $150$ steps with a batch size of $256$. We tune the learning rate in $\{0.01, 0.003, 0.001\}$. We use an early stopping strategy to stop training if the validation results do not improve for $20$ consecutive epochs.

We implement our method using HuggingFace transformers~\cite{wolf2020transformers} and FAISS library~\cite{douze2024faiss}. To ensure a fair comparison with the baseline that has the largest number of parameters (TIGER, 13M), we use a $2$-layer Transformer decoder with an embedding dimension of \( d = 448 \), a feed-forward layer dimension of \( 1024 \), and $4$ attention heads, which achieve a similar number of parameters. We tune the temperature $\tau \in \{0.03, 0.05, 0.07\}$ and semantic ID length $m\in \{4, 8, 16, 32, 64\}$. For the semantic encoder, we mainly use OpenAI's \texttt{text-embedding-3-large}\footnote{https://openai.com/index/new-embedding-models-and-api-updates/} to show the expressive abilities of long semantic IDs, but also conduct additional experiments using \texttt{sentence-t5-base}~\cite{ni2022sentence_t5} for fair comparison with benchmark results (\Cref{subsubsec:exp-expressive}). For model inference, we use a beam size $b=10$, tune the number of edges per node $k\in\{50, 100, 500\}$ and iteration steps $q\in \{2, 3, 5\}$. Please refer to~\Cref{app:details} for more details.

\subsection{Overall Performance}
We compare \mymodel\ with item ID-based and semantic ID-based baselines across three benchmarks and one larger public dataset. The results are shown in~\Cref{tab:overall}.

Across all datasets, methods leveraging semantic IDs generally outperform traditional item ID-based approaches. This aligns with the broader observation that incorporating item features and richer semantic information boosts recommendation quality. Within the semantic ID-based category, TIGER and HSTU consistently secure the second-best positions. These strong performances show the benefits of integrating semantic tokens into learning objectives over traditional methods.

Compared to all baselines, the proposed \mymodel\ achieves the best overall performance, ranking first in 11 out of 12 metrics. It outperforms the strongest baseline by an average of $12.6\%$ on the NDCG@10 metric. Similar to generative models, \mymodel\ uses semantic tokens as predictive targets, learning sub-item semantics. By generating tokens in semantic IDs in parallel, \mymodel\ efficiently and effectively leverages long and expressive semantic IDs with up to $64$ digits per item.

\subsection{Inference Efficiency Analysis}\label{subsubsec:exp-efficiency}

\begin{figure}[!t]
  \begin{center}
  \includegraphics[width=\columnwidth]{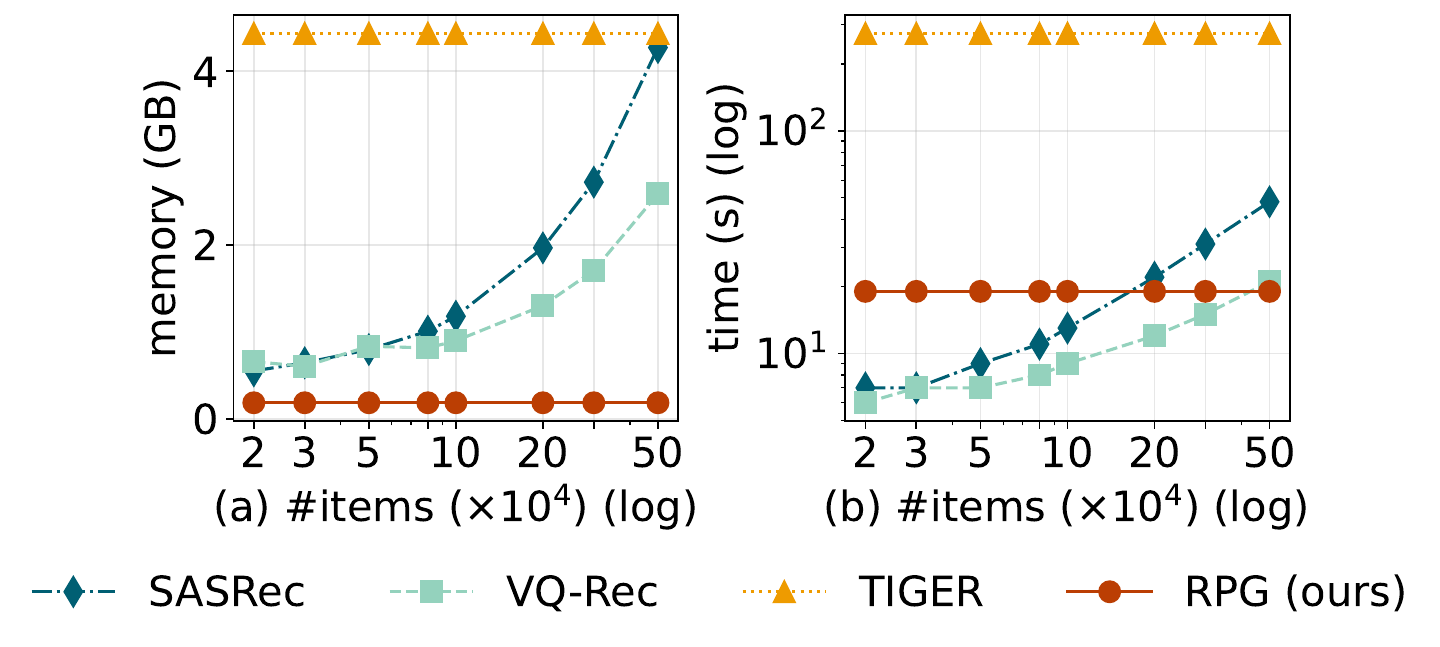}
  \caption{Comparison of runtime memory consumption and inference time \wrt the item pool size on the ``Sports'' dataset.}
  \label{fig:efficiency}
  \end{center}
\end{figure}

One of the main advantages of \mymodel\ is the efficient inference method (\Cref{sec:decoding}) that has a complexity unrelated to the size of the item pool. We've conducted efficiency analyses from a complexity perspective in~\Cref{subsubsec:complexity}. In this section, we empirically demonstrate the efficiency of RPG. Based on the ``Sports'' dataset, we add varying numbers of dummy items to the item pool, achieving item pool sizes from \(2 \times 10^4\) to \(50 \times 10^4\) items. We then measure the required runtime memory (GB) and clock time (s) for model inference over one epoch on the test set. We illustrate the results in~\Cref{fig:efficiency}. Note that all results are measured under a fixed set of hyperparameters, so the reported inference memory and time may not correspond to the best-performing configuration for each model. Here, runtime memory primarily refers to the GPU memory used to compute logits and generate the next tokens, including intermediate activations, rather than the total storage required to hold model parameters. A detailed breakdown of memory usage is provided in~\Cref{tab:mem-breakdown}.

We observe that both retrieval-based baselines, SASRec and VQ-Rec, require increasing memory and inference time as the item pool expands. In contrast, the generative baseline TIGER and the proposed \mymodel\ maintain constant memory usage and inference time regardless of the total number of items. Moreover, \mymodel\ reduces runtime memory consumption by nearly $25\times$ and achieves an inference speedup of almost $15\times$ compared to TIGER. These results show that \mymodel\ inherits the scalability of generative models (\ie inference cost independent of the number of items), while being significantly more efficient than existing generative baselines in both memory and time.

\begin{table}[t!]
  \small
  \centering
\caption{Ablation analysis of \mymodel. The recommendation performance is measured using NDCG@$10$. The best performance is denoted in \textbf{bold} fonts.}
\label{tab:ablation}
  \begin{tabular}{lcccc}
\toprule
\multicolumn{1}{c}{\textbf{Variants}} & \textbf{Sports} & \textbf{Beauty} & \textbf{Toys} & \textbf{CDs} \\
\midrule
\midrule
  \multicolumn{5}{@{}c}{\textit{Semantic ID Setting}} \\
  \midrule
  (1.1) OPQ $\rightarrow$ Random & 0.0179 & 0.0359 & 0.0288 & 0.0078 \\
  (1.2) OPQ $\rightarrow$ RQ & 0.0242 & 0.0421 & 0.0458 & 0.0406 \\
  \midrule
  \multicolumn{5}{@{}c}{\textit{Model Architecture}} \\
  \midrule
  (2.1) no proj. head & 0.0252 & 0.0423 & 0.0430 & 0.0361 \\
  (2.2) shared proj. head & 0.0256 & 0.0424 & 0.0438 & 0.0368 \\
  \midrule
  \multicolumn{5}{@{}c}{\textit{Model Inference}} \\
  \midrule
  (3.1) beam search & 0.0000 & 0.0000 & 0.0000 & 0.0000 \\
  (3.2) \emph{w/o} graph constraints & 0.0082 & 0.0214 & 0.0205 & 0.0183 \\
  \midrule
  \mymodel\ (ours) & \textbf{0.0263} & \textbf{0.0464} & \textbf{0.0490} & \textbf{0.0415} \\
  \bottomrule
\end{tabular}
\end{table}

\subsection{Ablation Study}

We conduct ablation analyses in~\Cref{tab:ablation} to study how each module contributes to the overall performance of \mymodel.

\noindent \hspace*{3mm} (1) We first examine whether OPQ is a good choice for tokenizing items into long semantic IDs. We replace OPQ with (1.1) randomly assigned discrete tokens or (1.2) RQ-based semantic IDs, and in both cases, performance drops. The drop is especially significant when using random tokens, indicating that the multi-token prediction loss effectively learns the semantics embedded in the long semantic IDs and fails when there is no semantic meaning in the IDs.

\noindent \hspace*{3mm} (2) We then examine the designs of the projection head used in the MTP loss (\Cref{eq:token-level-loss}). By default, we use a separate projection head for each digit of the semantic IDs. We implement two other variants: (2.1) removing the projection head and directly using the final hidden state of the sequence model to compute token logits, and (2.2) using a single shared projection head for all digits of the semantic IDs. Both variants lead to performance drops, highlighting the importance of mapping the sequence representation to different semantic spaces for decoding each digit of the tokens.

\noindent \hspace*{3mm} (3) We further experiment with two alternative inference methods: (3.1) beam search, which is commonly used in autoregressive generative models, and (3.2) inference without graph constraints, where the decoding graph is removed, and a similar number of semantic IDs are visited as in \mymodel, approximately 25\% of the item pool (\Cref{tab:reproduction}).
Our results show that beam search struggles because OPQ-based semantic ID lacks sequential dependencies. As a result, it easily falls into local optima during long beam search steps. Additionally, removing the decoding graph significantly reduces performance, highlighting its importance in guiding the generation of long semantic IDs.

\begin{figure}[!t]
  \begin{center}
  \includegraphics[width=\columnwidth]{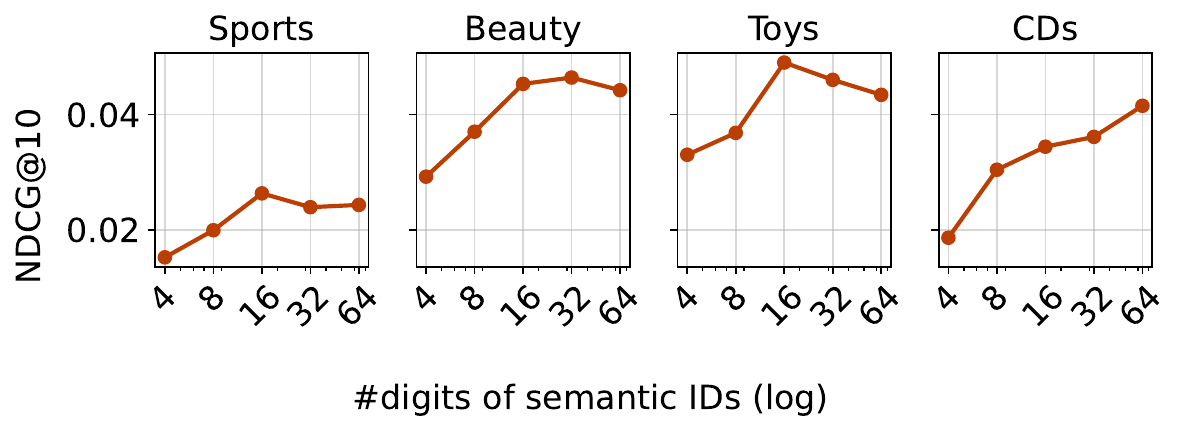}
  \caption{Scalability analysis of the performance (NDCG@10) of RPG \wrt semantic ID length.}
  \label{fig:scalability}
  \end{center}
\end{figure}

\subsection{Further Analysis}

\subsubsection{Scalability of Semantic ID Lengths}

The length of semantic IDs, \( m \) (\ie the number of digits in semantic IDs), is an important factor affecting recommendation efficiency and performance. We vary \( m \in \{4, 8, 16, 32, 64\} \) to examine how well the proposed method, \mymodel, scales with semantic ID length.  From~\Cref{fig:scalability}, we observe that, in general, longer OPQ-based semantic IDs lead to better performance. However, for smaller datasets, they may converge earlier. The optimal length for the ``Sports,'' ``Beauty,'' and ``Toys'' datasets is 16, 32, and 16, respectively. For our largest dataset ``CDs'', a semantic ID length of 64 achieves the best performance. These results demonstrate that our method scales effectively with increasing semantic ID length, particularly for larger datasets.

\subsubsection{Expressive Ability Analysis}\label{subsubsec:exp-expressive}

One assumption of using long semantic IDs is that they offer better expressive power than short ones. The original paper of TIGER indicates that they employ a Sentence-T5 semantic encoder (specifically, \texttt{sentence-t5-base} in our reproduction). In this experiment, we introduce a more powerful semantic encoder, OpenAI's \texttt{text-embedding-3-large}, to examine whether the semantic ID design can be expressive and benefit from the improved semantic encoder. We primarily compare TIGER with our proposed method, \mymodel, using different semantic encoders and varying semantic ID lengths.

From~\Cref{tab:expressive}, we observe that replacing \texttt{sentence-t5-base} with \texttt{text-embedding-3-large} does not always improve TIGER's performance when using 4-digit RQ-based semantic IDs. However, for \mymodel, a longer semantic ID performs better than a 4-digit one, even when using \texttt{sentence-t5-base}, and it already surpasses TIGER with the same encoder. Improving the encoder while maintaining long semantic IDs (up to 64 digits) further boosts performance. This demonstrates that the long semantic ID designed in \mymodel\ is more expressive and can benefit from a stronger semantic encoder.

\begin{table}[t!]
  \small
  \centering
\caption{Analysis of the expressive ability of short and long semantic IDs. The recommendation performance is measured using NDCG@$10$. ``PLM'' denotes the pretrained language model that encodes raw item features into representations.
  The best performance of each model is denoted in \textbf{bold}.
  }
\label{tab:expressive}
\setlength{\tabcolsep}{1mm}{
\resizebox{1\columnwidth}{!}{
  \begin{tabular}{ccccccc}
\toprule
\textbf{Model} & \textbf{PLM} & \textbf{\#digits} & \textbf{Sports} & \textbf{Beauty} & \textbf{Toys} & \textbf{CDs} \\
\midrule
\midrule
  \multirow{2}{*}{\textbf{TIGER}} & \texttt{sentence-t5-base} & 4 & 0.0225 & 0.0384 & \textbf{0.0432} & \textbf{0.0411} \\
  & \texttt{text-emb-3-large} & 4 & \textbf{0.0243} & \textbf{0.0411} & 0.0390 & 0.0409 \\
  \cmidrule(lr){1-3}
  \multirow{4}{*}{\textbf{\mymodel}} & \texttt{sentence-t5-base} & 4 & 0.0152 & 0.0292 & 0.0330 & 0.0186 \\
  & \texttt{text-emb-3-large} & 4 & 0.0117 & 0.0235 & 0.0275 & 0.0175 \\
   & \texttt{sentence-t5-base} & 16 \textasciitilde\ 64 & 0.0238 & 0.0429 & 0.0460 & 0.0380 \\
  & \texttt{text-emb-3-large} & 16 \textasciitilde\ 64 & \textbf{0.0263} & \textbf{0.0464} & \textbf{0.0490} & \textbf{0.0415} \\
  \bottomrule
\end{tabular}
}}
\end{table}

\begin{figure}[!t]
  \begin{center}
  \includegraphics[width=\columnwidth]{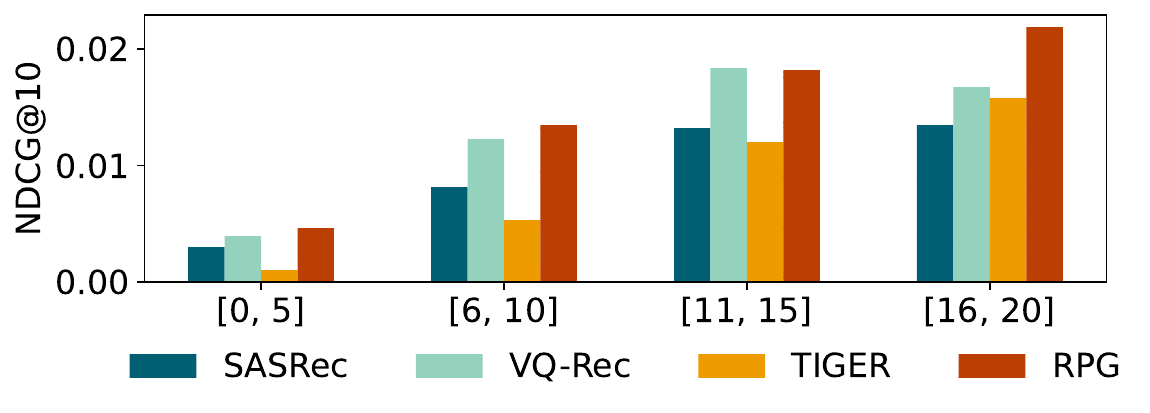}
  \caption{Cold-start analysis on the ``Sports'' dataset. Test items are divided into four groups based on the number of times they appear in the training set.
  }
  \label{fig:cold-start}
  \end{center}
\end{figure}

\subsubsection{Cold-Start Recommendation}

Since \mymodel\ optimizes by directly predicting semantic IDs within the token space, it is natural to hypothesize that it has a similar cold-start recommendation ability to other semantic ID-based methods like VQ-Rec and TIGER. We conduct experiments to evaluate how well \mymodel\ recommends long-tail items. We divide the test set items into four groups based on their occurrence in the training set: \texttt{[0, 5]}, \texttt{[6, 10]}, \texttt{[11, 15]}, and \texttt{[16, 20]}. As shown in~\Cref{fig:cold-start}, \mymodel\ achieves the best overall cold-start recommendation performance, demonstrating the strong ability of the MTP objective to learn item semantics. The long semantic IDs also enhance the tokens' semantic richness, further improving performance.

\subsubsection{Hyperparameter Analysis}

\begin{figure}[!t]
  \begin{center}
  \includegraphics[width=\columnwidth]{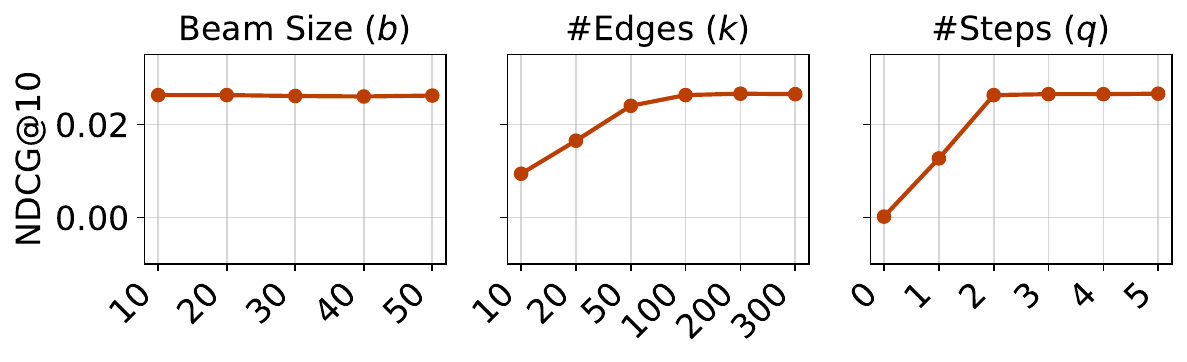}
  \caption{Model inference hyperparameter analysis. Each subplot shows the effect of varying one hyperparameter while keeping the others fixed.
  }
  \label{fig:param}
  \end{center}
\end{figure}

In this section, we tune the hyperparameters of the graph-constrained decoding method to analyze their impact on recommendation performance. The experiments are conducted on the ``Sports'' dataset with a default beam size of $b=10$, number of edges per node $k=50$, and iteration steps $q=2$. The results are shown in~\Cref{fig:param}.

\paratitle{Beam size $b$.} The beam size determines the number of semantic IDs retained in the beam at each step. In traditional beam search, a larger beam size brings the inference results closer to the global optimum, while a smaller one may lead to a local optimum. We vary the beam size as \( b \in \{10, 20, 30, 40, 50\} \). From~\Cref{fig:param}, we observe that, using NDCG@10 as the recommendation performance metric, a beam size larger than 10 does not differ significantly from 10. These results suggest that it is generally safe to use a relatively small beam size (as long as it is larger than the \( K \) in the top-\( K \) metric) to improve efficiency.

\paratitle{Number of edges per node $k$.}  The parameter \( k \) controls the number of edges per node in the graph, determining how many candidate connections are considered at each decoding step. A higher \( k \) allows the model to explore more diverse paths, potentially improving recommendation performance. We evaluate \( k \) with values \( \{10, 20, 50, 100, 200, 300\} \). From~\Cref{fig:param}, we observe a clear upward trend in NDCG@10 as \( k \) increases, with performance stabilizing beyond \( k = 100 \). This suggests that incorporating more edges enhances the model's ability to identify relevant recommendations, but excessive values provide diminishing returns. Since increasing $k$ also increases computation, we need to balance performance and efficiency by choosing an appropriate $k$.

\paratitle{Number of iteration steps $q$.}  The iteration steps \( q \) determine how many rounds of graph propagation are performed. We evaluate \( q \) with values \( \{0, 1, 2, 3, 4, 5\} \). As shown in~\Cref{fig:param}, increasing \( q \) significantly improves NDCG@10 from \( q=0 \) to \( q=2 \), indicating that additional iterations enhance recommendation performance. However, performance saturates after \( q=2 \), suggesting that the beam may converge for $q > 2$. This indicates that a small number of decoding steps (\eg \( q=2 \) or \( q=3 \)) is sufficient to reach optimal performance while maintaining efficiency.

\section{Related Work}

\paratitle{Sequential recommendation.} Sequential behavior patterns are useful for identifying user needs and providing recommendations. The sequential recommendation task usually takes a sequence of items as input. Early methods use Markov Chains to measure the item transitions~\cite{rendle2010fpmc}. 
In recent years, mainstream sequential recommender systems have come to use neural network models, including recurrent neural networks (RNN)~\cite{hidasi2016gru4rec,li2017narm,yue2024lrurec}, convolutional neural networks (CNN)~\cite{tang2018caser}, Transformers~\cite{kang2018sasrec,sun2019bert4rec}, and graph neural networks (GNN)~\cite{wu2019srgnn,chang2021surge}. 
Previous research mainly uses item IDs as the fundamental input of the sequential recommendation models. With the development of pretrained models and advanced feature encoders~\cite{liu2019roberta,devlin2019bert}, sequential recommendation models that are based on dense item features, usually called modality-based methods~\cite{hou2022unisrec,li2023recformer,yuan2023morec}, have also grown popular.
While using only modality features often makes optimization difficult~\cite{hou2022unisrec,hou2023vqrec}, researchers have addressed this by converting modality features into discrete tokens, known as semantic IDs~\cite{rajput2023tiger}. This approach has led to the development of both item-level models~\cite{petrov2024recjpq,hou2023vqrec} and token-level models~\cite{rajput2023tiger,zheng2024lcrec,zhai2024hstu} for the sequential recommendation task.

\paratitle{Semantic ID-based recommendation.} 
Semantic ID refers to a sequence of discrete tokens that jointly index an item while retaining some semantics~\cite{rajput2023tiger,hou2023vqrec,zheng2024lcrec,singh2024spmsid}.
Existing methods for constructing semantic IDs tokenize item features such as text~\cite{zheng2024lcrec,jin2024lmindexer}, images~\cite{liu2024mmgrec}, and collaborative signals~\cite{petrov2023gptrec,wang2024eager,wang2024colarec,liu2024mbgen} using techniques like quantization~\cite{rajput2023tiger,wang2024letter,zhu2024cost,zheng2025universal,zheng2025pre}, hierarchical clustering~\cite{hua2023p5cid,si2024seater}, or joint training with recommendation models~\cite{liu2024etegrec}.
By indexing items with semantic IDs, one can maintain a fixed vocabulary of tokens independent of the number of items, enabling scalable~\cite{zhai2024hstu,liu2024mbgen} and memory-efficient models~\cite{rajput2023tiger,tay2022dsi,petrov2024recjpq}.

Early semantic ID-based recommendation models were primarily retrieval-based. Each item's semantic ID was embedded, and the closest aggregated semantic ID embeddings to the user representation were retrieved~\cite{hou2023vqrec,petrov2024recjpq,yang2024liger}.
With the rise of generative models and insights from scaling laws in related fields, such as large language models (LLMs)~\cite{zhao2023llm_survey} and generative retrieval~\cite{tay2022dsi,li2024grsurvey,li2024grsurvey2}, generative recommendation has emerged as a promising paradigm for sequential recommendation~\cite{rajput2023tiger,zhai2024hstu,lin2025order,hou2025actionpiece,hou2025generative}.
In generative recommendation, tokens in semantic IDs directly serve as input for the sequence encoder. Models are then trained to autoregressively generate the tokens, which can be then parsed as predicted items.
These generative models explicitly integrate semantics into the learning objective and align more naturally with LLMs~\cite{zheng2024lcrec,jin2024lmindexer,tan2024idgenrec,li2025semantic}.
However, generative recommendation methods suffer from high inference latency due to their autoregressive nature.
Several studies have aimed to accelerate generative recommendation inference~\cite{lin2024efficient,ding2024specgr}, mainly by adapting techniques like speculative decoding to the recommendation task. However, these approaches do not fundamentally change the way semantic IDs are generated.
In this work, we explore an efficient semantic ID prediction approach that predicts each digit of a semantic ID in parallel.

\section{Conclusion}

In this work, we propose \mymodel, an efficient and effective semantic ID-based recommendation approach. Instead of autoregressively generating tokens, our key idea is to predict each digit in parallel. This removes the sequential dependency between digits, allowing the use of unordered, long, and expressive semantic IDs.
To enable parallel generation, we optimize the model with a multi-token prediction loss, explicitly integrating semantics into the learning objective. For inference, we leverage the observation that similar semantic IDs produce similar prediction logits. Based on this, we introduce a graph-constrained decoding method. By connecting semantic IDs with similar meanings, we iteratively refine the maintained beam of semantic IDs through graph propagation.
Experimental results show that \mymodel\ is both effective and efficient in terms of memory consumption and inference speed. For future work, we plan to explore how to align MTP-optimized semantic ID-based recommendation models with LLMs. Our goal is to develop an efficient LLM-based recommendation model that supports long semantic IDs with strong expressive ability.

\clearpage

\bibliographystyle{ACM-Reference-Format}
\balance
\bibliography{reference}

\appendix

\begin{center}
    {\Large \textbf{Appendices}}
\end{center}

\section{Notations}

We summarize the notations used throughout the paper in~\Cref{tab:notation}.

\section{Additional Implementation Details}\label{app:details}

All experiments were conducted on a single NVIDIA RTX 3090 GPU (24GB). For moderate-scale datasets (Sports, Beauty, Toys), each model training under a fixed hyperparameter setting takes less than 2 GPU hours. We tuned ``learning rate'', ``temperature'', and ``\#digit'', totaling 45 hyperparameter settings, resulting in under 90 GPU hours per dataset. For the large-scale dataset (CDs), each training takes under 8 hours, totaling about 180 GPU hours for all checkpoints reported. Regarding model inference, we evaluate all checkpoints on validation sets using the method described in~\Cref{subsubsec:logit}. After identifying the best checkpoint, we further tune inference-time hyperparameters ($k$ and $q$) over 9 combinations. This tuning process is lightweight and requires less than 1 GPU hour in total. We present the best hyperparameters for each dataset of \mymodel\ in~\Cref{tab:reproduction}.

\begin{table}[!t] %
    \small
	\caption{Notations and explanations.}
	\label{tab:notation}
	\vskip 0.1in
	\begin{tabular}{cl}
		\toprule
		\textbf{Notation} & \textbf{Explaination}\\
		\midrule
		$i_1$ & The item IDs \\
    $N$ & The number of items \\
    $t$ & The length of interaction sequences \\
    $c_1$ & The individual tokens in semantic IDs \\
    $m$ & The length (number of tokens) of semantic IDs \\
    $\mathcal{C}^{(1)}$ & The vocabulary of tokens in semantic IDs \\
    $M$ & The codebook size for each digit of the semantic ID \\
    $\bm{E}_j$ & The embedding lookup table for semantic IDs' $j$th digit \\
    $d$ & The dimension for token and item embeddings \\
    $\bm{e}_{1, c_1}$ & The token embedding \\
    $\bm{v}_i$ & The item representation \\
    $\bm{s}$ & The user (sequence) representation \\
    $\tau$ & The temperature hyperparameter \\
    $\operatorname{g}_j(\cdot)$ & The projection head \\
    $\bm{p}^{(j)}$ & The log probability (\ie logit) of the next token \\
    $k$ & The number of edges for each node in the decoding graph \\
    $b$ & The beam size \\
    $q$ & The number of steps in the decoding process \\
		\bottomrule
	\end{tabular}
\end{table}

\begin{table}[!t]
\centering
\small
\caption{Best hyperparameters of \mymodel\ for each dataset.}
\label{tab:reproduction}
\begin{tabular}{lcccc}
\toprule
\textbf{Hyperparameter} & \textbf{Sports} & \textbf{Beauty} & \textbf{Toys} & \textbf{CDs} \\
\midrule
semantic ID length $m$ & 16 & 32 & 16 & 64 \\
learning rate & 0.003 & 0.01 & 0.003 & 0.001 \\
temperature $\tau$ & 0.03 & 0.03 & 0.03 & 0.03 \\
beam size $b$ & 10 & 10 & 10 & 10 \\
\# edges per node $k$ & 100 & 100 & 50 & 500 \\
\# iteration steps $q$ & 2 & 3 & 5 & 3 \\
\midrule
\multicolumn{5}{@{}c}{\textit{Number/ratio of visited items under the best hyperparameter.}} \\
\midrule
\# visited items & $2\times 10^3$& $3\times 10^3$& $2.5\times 10^3$& $1.5\times 10^4$ \\
ratio of visited items & 10.90\% & 24.79\% & 20.97\% & 23.28\% \\
\bottomrule
\end{tabular}
\end{table}

\begin{table}[t!]
  \small
    \centering
    \caption{Performance and inference time per epoch of TIGER \wrt \#digit of semantic IDs. ``OOM'' denotes out-of-memory.}
    \label{tab:tiger-long-sid}
    \begin{tabular}{@{}rrr@{}}
      \toprule
      \multicolumn{1}{c}{\textbf{\#digit}} & \textbf{NDCG@10} & \textbf{Time (s)} \\
      \midrule
      \midrule
      4 & 0.0243 & 259 \\
      8 & 0.0219 & 788 \\
      16 & 0.0054 & 2,544 \\
      32 & OOM & OOM \\
      \bottomrule
    \end{tabular}
  \end{table}

\begin{table}[t!]
\small
  \centering
  \caption{Memory consumption breakdown. The term ``fetched storage'' refers to the amount of storage actually accessed to generate a ranking list for a user. The memory consumption of sequence encoders (\ie Transformer blocks) is omitted.}
  \label{tab:mem-breakdown}
  \begin{tabular}{@{}cccc@{}}
    \toprule
    \multicolumn{1}{c}{\textbf{Method}} & \textbf{NDCG@10} & \textbf{Storage} & \textbf{Fetched Storage} \\
    \midrule
    \midrule
    SASRec & 0.0192 & $O(Nd)$ & $O(\bm{\textcolor{blue}{N}}d)$ \\
    RecJPQ & 0.0102 & $O(Md + Nm)$ & $O(Md + \bm{\textcolor{blue}{N}}m)$ \\
    VQ-Rec & 0.0173 & $O(Mmd + Nm)$ & $O(Mmd + \bm{\textcolor{blue}{N}}m)$ \\
    RPG & 0.0263 & $O(Mmd + N(m+k))$ & $O(Mmd + \bm{\textcolor{blue}{bqk}}m)$ \\
    \bottomrule
  \end{tabular}
\end{table}

\section{Additional Discussion}

\textbf{What happens when equipping TIGER with long semantic IDs?}
To explore how existing semantic ID-based generative recommendation models perform with longer semantic IDs, we conduct experiments on the Sports dataset using TIGER~\cite{rajput2023tiger} as the base model. We employ \texttt{text-embedding-3-large} as the PLM to generate semantic IDs and increase the number of digits by applying more levels of residual quantization. As shown in~\Cref{tab:tiger-long-sid}, increasing the number of digits from $4$ to $16$ consistently degrades TIGER's performance and significantly slows down inference. These results highlight the difficulty of scaling semantic ID length in existing models. In contrast, RPG not only handles long semantic IDs efficiently (\Cref{fig:efficiency}) but also benefits from increased ID length in terms of performance (\Cref{fig:scalability}).

\noindent \textbf{Can multiple items share the same token set?}
Yes, this is possible. In RPG, each node in the decoding graph is initialized per item, rather than per unique token set. That said, since RPG uses long semantic IDs (\eg 64 tokens), the likelihood of two distinct items sharing the exact same token set is very low.

\noindent \textbf{Why does RPG perform worse than TIGER when \#digits = 4?}
As shown in~\Cref{tab:expressive}, RPG underperforms TIGER when both use semantic IDs of length $4$. However, comparing the two models under the same ID length is not particularly meaningful, given their fundamental differences in learning objectives, inference mechanisms, and semantic ID construction.
When using their optimal semantic ID lengths, RPG consistently outperforms TIGER across all datasets while also achieving better inference efficiency (\Cref{fig:efficiency}).

\noindent \textbf{Memory consumption breakdown.}
To better understand the memory footprint of RPG compared to existing methods, we provide a detailed breakdown in~\Cref{tab:mem-breakdown}. RPG consumes more total storage, as it maintains a token embedding table, an item-token mapping table, and a decoding graph. However, during inference, RPG is the only model that avoids accessing the entire storage but accesses a small subset of items, typically around 10-25\%.

\end{document}